\documentclass[pre,amsmath,onecolumn,amssymb,nofootinbib]{revtex4}
\usepackage{epsfig}
\usepackage{graphicx}
\usepackage{dcolumn}
\usepackage{bm}
\usepackage[english,russian]{babel}

\begin{document}

\title{Relaxation Processes in Many Particle Systems -- Recurrence Relations Approach}

\author{Anatolii V. Mokshin}
\email{anatolii.mokshin@mail.ru}
\address{Kazan Federal University,
Kazan, 420008, Russia}


\begin{abstract}
The general scheme for the
treatment of relaxation processes and temporal autocorrelations of
dynamical variables for many particle systems is presented in
framework of the recurrence relations approach. The time
autocorrelation functions and/or their spectral characteristics,
which are measurable experimentally (for example, due to
spectroscopy techniques) and accessible from particle dynamics
simulations, can be found by means of this approach, the main idea
of which is the estimation of the so-called frequency parameters.
Model cases with the exact and approximative solutions are given
and discussed.
\end{abstract}

\maketitle

\renewcommand{\baselinestretch}{1}
\normalsize

\section{Introduction} \noindent
Relaxation processes, which emerge in many particle systems, are
characterized by highly nontrivial features even for the cases of
the well-known simplified models~\cite{Zwanzig_2001}. So, for
example, the ideal gas dynamics at the drive by  external fields
exhibit the non-Markovian (memory) effects~\cite{Mokshin_PRL_2005}
as well as the manifestations of anomalous
transport~\cite{Uchaikin_UFN_2003}, whilst the chain of coupled
harmonic oscillators can display the nonlinear
dynamics~\cite{Risken_book_1989} with stochastic resonance
peculiarities~\cite{Hanggi_RMP_1998}. At the presence of
complicated fields of interactions in many particle systems
together with structural disorder and intricate spatiotemporal
correlations allows one to recognize these as the complex systems.
Thus, dense liquids, structural and spin glasses, foams, emulsions
and colloidal gels are the typical examples of the physical
complex systems, which combine the complicated dynamics together
with the structural inhomogeneity~\cite{Bertheir_2011}.

From theoretical standpoint, the description of many particle
dynamics reduces oneself into a unified fashion at the applying
the mathematical language of the distributions, the correlation
and relaxation functions, as well as the Green functions, that
provide a statistical treatment to some extent. Berne and Harp
marked the significance of time correlation functions in the
consideration of dynamic processes by the
phrase~\cite{Berne_Adv_Chem_Phys}: ``\ldots \emph{time correlation
functions have done for the theory of time-dependent processes
what the partitions functions have done for the equilibrium
theory. The time-dependent problem has became well defined}
\ldots''. These enthusiastic words become more clear and accepted,
if one takes into account that the correlation functions appear to
be directly related with the experimentally measured quantities
due to Kubo's linear response theory~\cite{Kubo_1957} as well as
by means of nonlinear response approach~\cite{Zausch_JPCM_2008},
which obtained recently a rapid development. Importantly, the time
correlation functions are associated with the concrete relaxation
processes and, thereby, provide the information about the proper
relaxation time scales~\cite{Mokshin_PRL_2005}. Moreover, these
functions can be applied to estimate quantitatively and simply the
so-called \textit{memory effects} in many particle system
dynamics~\cite{Mokshin_PRL_2005}, the dynamical heterogeneity
effects in particle movements~\cite{Rahman_PR_1964} and the
breakdown of system ergodicity~\cite{Lee_PRL_2007}.

Historically, the formulation of fluctuation-dissipation
theorem~\cite{Callen_PR_1951} and the Zwanzig-Mori's projection
operator formalism~\cite{Zwanzig_2001} can be distinguished as the
important milestones in development of the theoretical description
within time correlation and relaxation functions concept. If the
above-mentioned theorem was a forthright indication on the
relation between a (system) response function to an external field
and the fluctuations of the corresponding degree of freedom, then
the projection operator formalism was the first one, which had
provided us by a general scheme to define the time correlation
functions at the consideration of many particle system dynamics
within unified mathematical constrains. The projection operator
technique allowed one to establish the \textit{exact} relations
between characteristics of the different relaxation processes,
that had formed later a basis for the construction of such
theories like the generalized
hydrodynamics~\cite{Hansen/McDonald_book_2006} and the
mode-coupling
theories~\cite{Gotze_book_2009,Mokshin_Khusnutdinoff_TMF_2012}.
Despite obvious advances of the Zwanzig-Mori's projection operator
formalism in the description of many particle system dynamics, the
universal mathematical patterns in these systems became more clear
with the appearance of the recurrence relations
approach~\cite{Balucani_Phys_Rep_2003,Lee_JMP_1983}. In
particular, this approach gave a clear idea that the
(dis)similarity of relaxation time scales and the structural
recursive nature -- structural hierarchy -- can be mathematically
taken into account by means of recurrence relations.

The aim of this paper is to show the general scheme how the
relaxation processes in many particle systems can be treated
within the recurrence relations approach. Section $2$ presents the
some typical relations between the experimentally measured terms
and the correlation functions, which describe relaxation processes
in many particle systems. The formulation of the basic general
relations is given in section $3$. The simplified model cases with
the exact solutions are presented in sections $4$ and $5$, whereas
the description of the density fluctuations in simple liquids is
considered in section $6$. And, finally, we conclude in Section
$7$.

\section{Correlation functions vs. observable terms}

\noindent Let us consider the system of $N$ interacted particles,
which evolves at the temperature $T$ within the volume $V$. The
full dynamics of the system is characterized by a set of the
variables $A_n(t)$ like, for instance, local density, particle
displacement, particle velocity, dipole moment etc. Nevertheless,
a concrete problem under study enforces, as a rule, to restrict
oneself by a some variable associated with phenomena. The choice
of the dynamical variable can be caused by the experimentally
measurable response function, which is related with the
corresponding relaxation (or correlation) function of this
dynamical variable.

\textit{Scattering techniques.} -- The inelastic neutron
scattering and the inelastic X-ray scattering techniques allow one
to measure the dynamic structure factor $S(k,\omega)$ and the
incoherent scattering function $S_s(k,\omega)$, where $k$ is the
wave number and $\omega$ is the frequency. These terms are related
with the autocorrelation functions of the spatial Fourier
transforms for the local density fluctuations
$\rho(\textbf{r},t)=(1/\sqrt{N})\sum_{i}
\delta[\mathbf{r}-\mathbf{r}_i(t)]$ and for the tagged (single)
particle displacement
$\rho_s(\textbf{r},t)=\delta[\mathbf{r}-\mathbf{r}_i(t)]$,
correspondingly,~\cite{Hansen/McDonald_book_2006}
\begin{equation} \label{eq: dynamic_structure_factor}
S(k,\omega) = \frac{S(k)}{2 \pi N} \int_{-\infty}^{\infty} dt\;
e^{-i\omega t} \sum_{i,j} \left . \frac{\langle
e^{-i\mathbf{k}\cdot \mathbf{r}_i(0)} e^{i\mathbf{k}\cdot
\mathbf{r}_j(t)} \rangle}{\langle e^{-i\mathbf{k}\cdot[
\mathbf{r}_i(0)-\mathbf{r}_j(0)]} \rangle} \right .
\end{equation}
\begin{equation}
S_s(k,\omega) = \frac{1}{2 \pi N} \int_{-\infty}^{\infty} dt\;
e^{-i\omega t} \sum_{i} \langle e^{-i\mathbf{k}\cdot
\mathbf{r}_i(0)} e^{i\mathbf{k}\cdot \mathbf{r}_i(t)} \rangle ,
\end{equation}
where $\mathbf{k}$ is the wave vector, $S(k) = (1/N) \sum_{i,j}
\left \langle e^{-i\mathbf{k}\cdot[
\mathbf{r}_i(0)-\mathbf{r}_j(0)]} \right \rangle =
\int_{-\infty}^{\infty} d\omega \; S(k,\omega)$ is the static
structure factor, $k=|\mathbf{k}|$ and $\rho(\textbf{k},t) =
(1/\sqrt{N}) \sum_{i}e^{i\mathbf{k}\cdot \mathbf{r}_i(t)}$ is the
space Fourier transform of $\rho(\textbf{r},t)$. Thus, the dynamic
structure factor $S(k,\omega)$ estimates the collective dynamics
with frequencies $\omega$ over spatial scales $\sim 2\pi/k$ and is
related with the density autocorrelation function (or the
so-called intermediate scattering function)
\begin{equation}
\phi_{coh}(k,t) = \frac{1}{N} \sum_{i,j} \left . \frac{\langle
e^{-i\mathbf{k}\cdot \mathbf{r}_i(0)} e^{i\mathbf{k}\cdot
\mathbf{r}_j(t)} \rangle}{\langle e^{-i\mathbf{k}\cdot[
\mathbf{r}_i(0)-\mathbf{r}_j(0)]} \rangle} \right . ,
\end{equation}
whereas the incoherent scattering function $S_s(k,\omega)$
performs the same for single particle dynamics and is associated
with the particle displacement autocorrelation function (the
self-intermediate scattering function)
\begin{equation}
\phi_{self}(k,t) = \frac{1}{N} \sum_{i} \left . \langle
e^{-i\mathbf{k}\cdot \mathbf{r}_i(0)} e^{i\mathbf{k}\cdot
\mathbf{r}_i(t)} \rangle \right . .
\end{equation}

\textit{Dielectric spectroscopy.}  --  The complex dielectric
permittivity $\varepsilon^*(\omega)$ is measurable due to
dielectric spectroscopy experiments. This term is related with the
macroscopic dipole correlation function
\begin{equation}
\phi_d(t) = \frac{\langle \mathbf{M}(0)\mathbf{M}(t)
\rangle}{\langle \mathbf{M}(0)^2 \rangle}
\end{equation}
by the relation \cite{Feldman_book_2006}
\begin{equation}
1 - \frac{\varepsilon^*(\omega) -
\varepsilon_{\infty}}{\varepsilon_s - \varepsilon_{\infty}} = \Im
\left [ s \int_0^{t} dt\; e^{-s t} \phi(t)\right ],\; s=i\omega,
\end{equation}
where $\mathbf{M}(t)$ is the macroscopic fluctuating dipole moment
of the sample volume unit, which is equal to the vector sum of all
the molecular dipoles, $\varepsilon_s$ and $\varepsilon_{\infty}$
are the low- and high-frequency limits of the dielectric
permittivity, respectively, and $\Im[\ldots]$ means imaginary part
of $[\ldots]$.

\textit{Transport coefficients.} -- A feature of resonance
techniques is that they can measure the transport properties. So,
for example, the nuclear magnetic resonance provides the
information about the self-diffusion coefficient $D_s$ and
viscosity $\eta$ of study system, which are (as well as other
transport coefficients) are related with the autocorrelation
functions from the corresponding current dynamical variables
through the Green-Kubo relations~\cite{Hansen/McDonald_book_2006}.
So, for self-diffusion coefficient one has
\begin{equation} \label{eq: self_diffusion}
D = \frac{k_B T}{m} \int_0^{\infty} dt \; \frac{\langle
\textbf{v}(0) \textbf{v}(t) \rangle}{\langle \textbf{v}(0)
\textbf{v}(0) \rangle},
\end{equation}
where
\begin{equation}
\phi_{vel} = \frac{\langle \textbf{v}(0) \textbf{v}(t)
\rangle}{\langle \textbf{v}(0) \textbf{v}(0) \rangle}
\end{equation}
is the single particle velocity autocorrelation function.
Equation, similar to Eq.~(\ref{eq: self_diffusion}), can be
written for rotational diffusion coefficient in term of the
dynamical variable -- angular velocity $\Omega_{\alpha}$, that
accounts the molecular reorientations measurable in the
depolarization of fluorescence studies.

Further, the shear viscosity is
\begin{equation} \label{eq: shear_viscosity}
\eta = \frac{1}{k_B T V} \int_0^{\infty} dt \; \left \langle
P_{xy}(0) P_{xy}(t) \right \rangle ,
\end{equation}
where
\begin{equation}
\phi_{\eta}(t) = \langle P_{xy}(0) P_{xy}(t) \rangle
\end{equation}
is the autocorrelation function of the components of the pressure
tensor, which are given by virial formula
\begin{equation}
P_{\alpha \beta} = \sum_{i=1}^{N} \left ( m v_{i\alpha}v_{i\beta}
+ \frac{1}{2} \sum_{i\neq j}^{N} F_{ij\alpha} r_{ij\beta} \right
), \; \; \alpha,\beta =x,y,z,
\end{equation}
and $F_{ij\alpha}$ denotes the $\alpha$-component of the force
between particles $i$ and $j$ that are at distance $r_{ij}$ from
one another.

The thermal conductivity $\lambda$ can be expressed as
\begin{equation} \label{eq: thermal_conductivity}
\lambda = \frac{1}{k_B T^2 V} \int_{0}^{\infty} dt\; \left \langle
J_0^{ez}(0)J_0^{ez}(t) \right \rangle,
\end{equation}
where
\begin{equation}
\phi_{\lambda} = \left \langle J_0^{ez}(0)J_0^{ez}(t) \right
\rangle
\end{equation}
is the heat current autocorrelation function with the dynamical
variable
\begin{equation}
J_0^{ez} = \sum_{i=1}^{N} v_{iz} \left ( \frac{m
|\textbf{v}_i|^2}{2} + \frac{1}{2}\sum_{i\neq j}^N U(r_{ij})
\right ) - \frac{1}{2} \sum_{i=1}^{N} \sum_{i\neq j}^N
\textbf{v}_i \textbf{r}_{ij}\frac{\partial U(r_{ij})}{\partial
z_{ij}},
\end{equation}
$U(r_{ij})$ is the potential of particle interaction.

The examples given above show clearly that a variety of quantities
experimentally observed can be treated in terms of the time
correlation (or relaxation) functions. Relation of other
quantities experimentally measured with the corresponding
correlation functions can be found, for example, in comprehensive
review \cite{Berne_Adv_Chem_Phys}. Thus, the problem of the
explanation of the experimental results in some (not all, but
many) cases can be reduced to the problem of the finding either
the proper relaxation function or, at least, the asymptotic
behavior of this function.

\section{Theoretical background}
\noindent Let us assume that we consider the dynamical variable
$A(t)$, the time evolution of which is defined by the Heisenberg
equation
\begin{equation} \label{eq: Heisenberg_eq}
\frac{dA(t)}{dt}= \mathrm{i}[H,A(t)] =
\mathrm{i}\hat{\mathcal{L}}A(t),\ \ \ \ A(t)|_{t=0}=A,
\end{equation}
where $H$ is the Hamiltonian of the system, $\hat{\mathcal{L}}$ is
the Liouville operator, which is taken to be Hermitian and $[\ .\
,\ .\ ]$ is the Poisson bracket. The formal solution of equation
(\ref{eq: Heisenberg_eq}) can be written as
\begin{equation}
A(t) = \mathrm{e}^{\mathrm{i}\hat{\mathcal{L}}t}A.
\end{equation}
On the other hand, the Hamiltonian $H$ defines the averaging
operation $\langle A \rangle$ through the density of phase space
$\rho \propto \exp{[-\beta(H-\mu N)]}$, where $\mu$ is the
chemical potential, $\beta=(k_BT)^{-1}$ and $k_B$ is the Boltzmann
constant. Then, in classical limit one has a simple correspondence
between the scalar product of a pair of dynamical variables $A$
and $B$ of the Liouville space and the corresponding correlation
function~\cite{Hansen/McDonald_book_2006}
\begin{equation} \label{eq: space_identity}
(A,B) \equiv \langle A B^* \rangle .
\end{equation}
The symbol $*$ marks the complex conjugation. In fact, equation
(\ref{eq: space_identity}) provides the identity between the
relaxation function $(A(t),B)$ and the time correlation function
$\langle A(t) B^* \rangle$, i.e. $(A(t),B) \equiv \langle A(t) B^*
\rangle$. Further, we will utilize the time autocorrelation
functions (TACF) in the dimensionless form
\begin{equation} \label{eq: TCF}
\phi(t) = \frac{\langle A(0)^* A(t)\rangle}{\langle
|A(0)|^2\rangle},
\end{equation}
that ensures the fulfillment of the next conditions
\begin{eqnarray}
\phi(t)|_{t=0}=1, \ \ \ 1\geq \phi(t) \geq 0,\ \ \ \left .
\frac{d\phi(t)}{dt}\right |_{t=0} = 0.
\end{eqnarray}
Such a representation of the TACF allows one to focus on the time
dependence of $\phi(t)$ directly, and to do the comparison of
autocorrelations for different dynamical variables.

Applying the Gram-Schmidt orthogonalization procedure at the
initial condition $A_0 \equiv A$, we generate the set of dynamical
variables:
\begin{subequations}
\begin{equation}
\mathbf{A} = \{A_0,\ A_1,\ A_2,\ldots , \ A_{\nu}, \ldots \},
\end{equation}
\begin{eqnarray}
& &(A_{\nu},A_{\mu}) = (A_{\nu},A_{\nu})\delta_{\nu,\mu},\\
& &\nu,\mu=0,\ 1,\ 2,\ 3,\ldots , \nonumber
\end{eqnarray}
\end{subequations}
which are related by the recurrence relation
\begin{subequations}
\begin{equation} \label{eq:recurrant_1}
A_{\nu+1} = \mathrm{i}\hat{\mathcal{L}}A_{\nu} +
\Delta_{\nu}A_{\nu-1},
\end{equation}
\begin{eqnarray}
\label{eq: freq_parameters}
\Delta_{\nu} = \frac{(A_{\nu},A_{\nu})}{(A_{\nu-1},A_{\nu-1})},\\
A_{-1}=0,\ \ \ \Delta_{0}=1,\nonumber
\end{eqnarray}
\end{subequations}
where $\Delta_{\nu}$ are the frequency parameters with the
dimension of the squared frequency, $\delta_{\nu,\mu}$ is the
Kronecker delta. From the definition (\ref{eq: freq_parameters})
one can see that the physical meaning of the parameters
$\Delta_{\nu}$ depends on the concrete process and is defined by
the corresponding dynamical variables $A_{\nu-1}$ and $A_{\nu}$.
Equation (\ref{eq:recurrant_1}) is also known from the recurrence
relations approach as \textit{the first recurrence
relation}~\cite{Lee_PRE_2000,Lee_PRE_2000_1,Lee_JMP_1983}.
Usually, it is convenient (but not necessarily) to perform the
construction of the set $\mathbf{A}$ on the basis of the dynamical
variable $A_0$, which is associated with the processes
experimentally studied.

Both the projection operators technique~\cite{Zwanzig_2001} as
well as the recurrence relations
approach~\cite{Lee_PRE_2000_1,Lee_PRE_2000} yield the chain of the
integro-differential equations for the dynamical variables
$\mathbf{A}$ of the form
\begin{eqnarray} \label{eq: GLE}
\frac{d}{dt}A_{\nu}(t) &=& - \Delta_{\nu+1} \int_{0}^{t}
A_{\nu}(t-\tau)\frac{\langle A_{\nu+1}(0)^* A_{\nu+1}(\tau)
\rangle}{\langle|A_{\nu+1}(0)|^2\rangle}d\tau + A_{\nu+1}(t),\\
& &\mathrm{here\ and\ hereafter}\ \nu = 0,\ 1,\ 2,\ 3,\ldots ,
\nonumber
\end{eqnarray}
that is the exact consequence of equation~(\ref{eq:
Heisenberg_eq}).\footnote{For a case, when the dynamical variable
is chosen to be the particle velocity, i.e. $A_0=v$,
equation~(\ref{eq: GLE}) represents the known generalized Langevin
equation.} For the TACF's defined as
\[
\phi_{\nu}(t)= \frac{\langle A_{\nu}(0)^* A_{\nu}(t)
\rangle}{\langle|A_{\nu}(0)|^2\rangle},
\]
the chain~(\ref{eq: GLE}) takes the following form
\begin{equation} \label{eq: chain}
\frac{d}{dt}\phi_{\nu}(t) =  - \Delta_{\nu+1} \int_{0}^{t}
\phi_{\nu}(t-\tau)\phi_{\nu+1}(\tau) d\tau.
\end{equation}
Then, applying the Laplace transform operator
\begin{equation} \label{eq: laplace_transfrom}
\hat{\mathcal{L}}[f(\tau)] = \tilde{f}(s) = \int_0^{\infty}
e^{-s\tau} f(\tau) d\tau
\end{equation}
to chain of equations~(\ref{eq: chain}), one obtain the recurrence
formula
\begin{equation} \label{eq: freq_chain}
\tilde{\phi}_{\nu}(s) = \frac{1}{s + \Delta_{\nu+1}
\tilde{\phi}_{\nu+1}(s)},
\end{equation}
which can be transformed to continued fraction representation of
frequency spectrum of the TACF:
\begin{equation} \label{eq: continued_fraction}
\tilde{\phi}_0(s) = \cfrac{1}{s+ \cfrac{\Delta_1}{s+
\cfrac{\Delta_2}{s+\cfrac{\Delta_3}{s+\ddots}}}}.
\end{equation}
On the one hand, equation (\ref{eq: continued_fraction}) indicates
that the form of the spectrum $\tilde{\phi}_0(s)$ and, thereby, of
the relaxation function $\phi_0(t)$, is completely defined by the
frequency parameters $\Delta_{\nu}$ as well as by the ratios
$\Delta_{\nu+1}/\Delta_{\nu}$. On the other hand, the values of
$\Delta_{\nu}$'s are directly associated with the frequency/time
range, for which the fraction solution [equation~(\ref{eq:
continued_fraction})] will be relevant. Thus, the problem of
finding $\phi_0(t)$ [or $\tilde{\phi}_0(s)$] is reduced,
mathematically, to the problem of finding a function, which is
representable in the form of continued fraction with the some
unique set of values of $\Delta_{\nu}$'s. Although these frequency
parameters are physical characteristics of the  concrete
relaxation processes [see equation~(\ref{eq: freq_parameters})],
one can possible to consider some different model situations,
which can be realized in some general cases.

\section{Models of the finite sets of variables}

\noindent Let us consider the cases with the finite set of
dynamical variables ($\nu$ is finite), that corresponds to the
finite-dimensional Liouville spaces. Such situations arise at the
condition with $A_{\nu}=0$ and $\Delta_{\nu}=0$; and are relevant
to nonergodic processes with non-decaying correlation functions
expressed by cosine functions.

\subsection{A case of $\nu=2$}

\noindent At the condition $\nu=2$ one has $A_2=0$ and
$\Delta_2=0$.\footnote{It is clear that the case of $\nu=1$ is
trivial. Therefore, this case is not considered here.} Then,
continued fraction (\ref{eq: continued_fraction}) yields the
system of two equations
\begin{equation}
\left\{
\begin{array}{c}
 1 - s\tilde{\phi_0}(s)  = \Delta_1 \tilde{\phi}_0(s)
\tilde{\phi}_1(s), \\
s \tilde{\phi}_1(s)= 1
\end{array} \right.
\end{equation}
with simple solutions in the time domain
\begin{subequations}
\begin{equation}
\phi_0(t)= \cos(\sqrt{\Delta_1}t),\label{eq: frr}
\end{equation}
\begin{equation}
\phi_1(t)= 1.
\end{equation}
\end{subequations}
The relaxation function of the form~(\ref{eq: frr}) reproduces the
behavior of undamped harmonic oscillator.  This situation is
realized, for example, in the case of density fluctuations of
homogeneous electron gas at the finite wave numbers $k$ and the
temperature $T=0$. Other example, where the case appears, is the
dynamics of the chain of classical harmonic oscillators. Here, the
TACF of particle velocity, $\phi_0(t) = \langle \upsilon(0)
\upsilon(t) \rangle/\langle \upsilon(0)^2 \rangle$, is described
by equation~(\ref{eq: frr}) (see reference
\cite{Florencio_Lee_1985}).

\subsection{A case of $\nu=3$}

\noindent One has here that $A_3=0$ and $\Delta_3=0$. Then,
continued fraction~(\ref{eq: continued_fraction}) in this case
transforms into the next system of equations:
\begin{equation}
\left\{
\begin{array}{c}
 1 - s\tilde{\phi_0}(s)  = \Delta_1 \tilde{\phi}_0(s)
\tilde{\phi}_1(s), \\
 1 - s\tilde{\phi_1}(s)  = \Delta_2 \tilde{\phi}_1(s)
\tilde{\phi}_2(s), \\
s \tilde{\phi}_2(s)= 1,
\end{array}
\right.
\end{equation}
which can be resolved by means of the inverse Laplace transform
$\hat{\mathcal{L}}^{-1}$ and yields the solutions
\begin{subequations}
\begin{equation} \label{eq: nu_2}
\phi_0(t)= \frac{1}{\Delta_1+ \Delta_2}\left [ \Delta_2 +\Delta_1
\cos(\sqrt{\Delta_1+\Delta_2}t) \right ],
\end{equation}
\begin{equation}
\phi_1(t)= \cos(\sqrt{\Delta_2}t).
\end{equation}
\begin{equation}
\phi_2(t)= 1.
\end{equation}
\end{subequations}
Equation~(\ref{eq: nu_2}) corresponds again to the harmonic
behavior of the initial TACF $\phi_0(t)$, where the period is
defined by two frequency parameters, $\Delta_1$ and $\Delta_2$.

From these two cases presented above, one can see that it is
possible to find exact analytical solutions for the relaxations
functions $\phi_{\nu}(t)$ at any finite dimension $\nu$.

\section{Models of the infinite sets of variables}

\noindent For the infinite-dimensional Liouville spaces, $\nu \to
\infty$, the set of possible scenarios for the TACF $\phi_0(t)$ is
vast and contains the decaying functions.

\subsection{Gaussian relaxation}
\noindent  Let us now consider the case, where the frequency
parameters $\Delta_{\nu}$'s are related according to the
arithmetic progression:
\begin{equation} \label{eq: gaussian_rec}
\Delta_1, \ \ \Delta_2 = 2 \Delta_1, \ \ \Delta_3 = 3\Delta_1,\ \
\ldots,\ \  \Delta_{\nu} = \nu\Delta_{1}.
\end{equation}
Then, the continued fraction~(\ref{eq: continued_fraction}) takes
the following form
\begin{equation} \label{eq: continued_fraction_Gaussian}
\tilde{\phi}_0(s) = \cfrac{1}{s+ \cfrac{\Delta_1}{s+
\cfrac{2\Delta_1}{s+\cfrac{3\Delta_1}{s+\ddots}}}},
\end{equation}
that corresponds in the time domain to the ordinary Gaussian
function~\cite{Abramowitz_1972}
\begin{equation} \label{eq: Gaussian}
\phi_0(t) = \mathrm{e}^{-\Delta_1 t^2/2}.
\end{equation}
The most well-known examples of the physical realization of such
relaxation is the density fluctuations in the perfect gas and
one-particle dynamics in liquids (at the limit of high wave
numbers $k$)~\cite{Mokshin_JPCM_2006,Hansen/McDonald_book_2006}.
Other case with this relaxation is the dynamics of the
one-dimensional XY-model at $T\to
\infty$~\cite{Florencio_Lee_1987}.

The exact correspondence between the frequency parameters [given
by relations~(\ref{eq: gaussian_rec})] indicates on the
possibility to study quantitatively the deviation from the
Gaussian relaxation by means of the simple comparison of the
parameters $\Delta_{\nu}$'s:
\begin{equation} \label{eq: non-Gaussian}
\alpha_{\nu} =
\frac{\nu}{\nu+1}\frac{\Delta_{\nu+1}}{\Delta_{\nu}}-1.
\end{equation}
For the Gaussian relaxation one has $\alpha_{\nu}=0$, whilst
deviations from the zeroth values of $\alpha_{\nu}$ will be caused
by manifestations of the non-Gaussian behavior of $\phi_0(t)$.

\subsection{Damped relaxation of oscillated correlator}
\noindent Let us consider the specific case, where the frequency
parameters are finite and equal to each other
\begin{equation} \label{eq: condition_for_Bessel}
\Delta_1 = \Delta_2 =  \Delta_3 =  \ldots = \Delta_{\nu}.
\end{equation}
that corresponds to the continued fraction~(\ref{eq:
continued_fraction}) of the form
\begin{equation} \label{eq: continued_fraction_Bessel}
\tilde{\phi}_0(s) = \cfrac{1}{s+ \cfrac{\Delta_1}{s+
\cfrac{\Delta_1}{s+\ddots}}}.
\end{equation}
As known from the theory of continued fractions,
expression~(\ref{eq: continued_fraction_Bessel}) is the
representation of the next function (over the variable $s$):
\begin{equation} \label{eq: Bessel_freq}
\tilde{\phi}_0(s)=\frac{-s+\sqrt{s^2+4\Delta_1}}{2\Delta_1}.
\end{equation}
Applying the inverse Laplace transform operator
$\hat{\mathcal{L}}^{-1}$ to equation~(\ref{eq: Bessel_freq}), one
obtains the TACF
\begin{equation} \label{eq: Bessel_time}
\phi_0(t) = \frac{1}{\sqrt{\Delta_1}t}J_{1}(2\sqrt{\Delta_1}t),
\end{equation}
where $J_1$ is the Bessel function of the first order. Such
relaxation appears in the processes, which characterized by the
damped harmonic oscillatory behavior. For example,
expression~(\ref{eq: Bessel_time}) is the exact form for the TACF
of velocity of the Brownian particle in linear chain of the
identical harmonic
oscillators~\cite{Rubin_PR_1963,Mokshin_PRL_2005,Zwanzig_2001}.
Moreover, the dynamics of the two-dimensional electron gas at the
temperature $T=0$ and at the defined range of the wave number $k$
is other manifestation of relaxation with the TACF of the
form~(\ref{eq: Bessel_time})~(see
reference~\cite{Lee_Physica_Scr_1987}).

Thus, the presented cases demonstrate that at the known
correspondence between the frequency parameters
$\mathcal{F}(\Delta_1,\Delta_2,\Delta_3,\ldots,\Delta_{\nu},\ldots)$,
one can \textit{exactly} define the initial TACF $\phi_0(t)$ and
to estimate the frequency features of its spectral image
$\tilde{\phi}_0(s)$.

\section{Density fluctuations in simple liquids}

\subsection{Frequency parameters}

\noindent Let us now consider the liquid system, where the
particle interactions are characterized by the spherical symmetry
and the potential contains the radial dependence only. Liquid
metals and condensate noble gases are the typical examples of such
systems~\cite{Mokshin_PRE_2001,Mokshin_JETP_Lett_2002,Hansen/McDonald_book_2006}.
Further, we take the space Fourier transform of the local density
fluctuations, $A_0(\mathbf{k}) = \rho(\textbf{k},t) = (1/\sqrt{N})
\sum_{i}e^{i\mathbf{k}\cdot \mathbf{r}_i(t)}$, as the initial
dynamical variable, the TACF of which, $\phi_{coh}(k,t)$, is
related with dynamic structure factor $S(k,\omega)$ [see
equation~(\ref{eq: dynamic_structure_factor})]. Then, the
frequency parameters can be found according to
relations~(\ref{eq:recurrant_1}) and (\ref{eq: freq_parameters})
\begin{subequations}
\begin{eqnarray}
(A_0(k),A_0(k)) &=& S(k),\ \ \ (A_1(k),A_1(k)) = \frac{k_BT}{m}
k^2,\nonumber\\
\Delta_1(k) &=& \frac{k_B T}{m} \frac{k^2}{S(k)} = \frac{(v_T
k)^2}{S(k)}, \label{eq: delta_1}
\end{eqnarray}
\begin{eqnarray} \label{eq: delta_2}
\Delta_2(k) = 3\frac{k_BT}{m}k^2 + \frac{\rho}{m} \int \nabla_l^2
u(r) [1- \exp(i\mathbf{k}\mathbf{r})]g(r)d^3\mathbf{r} - \Delta_1,
\end{eqnarray}
\begin{equation} \label{eq: delta_3}
\Delta_3(k) = \frac{1}{\Delta_2(k)} \Xi(k) - \frac{[\Delta_1(k) +
\Delta^2(k)]^2}{\Delta_2(k)},
\end{equation}
\end{subequations}
where $v_T$ is the average thermal velocity of particles, $\rho$
is the number density, $g(r)$ is the pair distribution function,
$u(r)$ is the interparticle potential, the suffix $l$ marks the
component parallel to wave vector $\mathbf{k}$. Within the
assumption about pair-additivity of the interparticle potential
$u(r)$, the term $\Xi(k)$ takes the following
form~\cite{Bansal_PRA_1974}:
\begin{eqnarray}
\Xi(k) &=& 15 \left ( \frac{k_{B}T}{m}\right )^{2}k^{4}
+\frac{k_{B}T}{m}k^{2} \frac{\rho}{m} \int d
\textbf{r} \; g(r)\nabla_{l}^{2}u(r) \\
&+& 6\rho\frac{k_{B}T}{m^{2}} k \int d^3\textbf{r} \; g(r)
\nabla_{l}^{3}u(r) \sin(\textbf{kr}) \nonumber\\
&+&2\frac{\rho}{m^{2}}\int d^3\textbf{r} \; g(r) [\nabla
\nabla_{l}u(r)]^{2} [1-\cos(\textbf{kr})] \nonumber\\
&+& \frac{\rho^2}{m^2} \int \int \; d^3\textbf{r} d^3\textbf{r}'
g_{3}(\textbf{r,r}') [1-\cos(\textbf{kr} -\textbf{kr}')]
\left(\nabla \nabla_{l} u(r) \right ) \left(\nabla ' \nabla_{l}'
u(r') \right ),\nonumber
\end{eqnarray}
where $g_3(\textbf{r},\textbf{r}')$ is the three-particle
distribution function. The frequency parameters of a higher order
will contain the more complicated integral expressions with the
$n$-particles distribution functions $g_{n}(\bar{r})$, i.e.
$\Delta_{\nu}(k) =
F[\Delta_1(k),\Delta_2(k),\ldots,\Delta_{\nu-1}(k);g(r),g_3(\bar{r}),\ldots,g_{\nu}(\bar{r})]$.

\subsection{Dynamics at high wave numbers and short time scales}

\noindent In the ranges of the wave numbers $k$ larger than
$k_m=2\pi/\sigma$  [where $\sigma$ is the effective particle size
and $k_m$ is the first maximum in the static structure factor
$S(k)$] and of the short time scales, which correspond to lengths
smaller than the mean free path, the interactions between
particles in simple liquids can be neglected. Taking into account
that for this range of $k$ one has $S(k) \to 0$, we obtain
directly from equations (\ref{eq: delta_1}), (\ref{eq: delta_2})
and (\ref{eq: delta_3})
\begin{equation} \label{eq: free_part_limit}
\Delta_1(k) = (v_Tk)^2, \ \ \ \Delta_2(k) = 2 (v_Tk)^2, \ \ \
\Delta_3(k) = 3 (v_Tk)^2.
\end{equation}
Comparing (\ref{eq: free_part_limit}) and (\ref{eq: gaussian_rec})
one can see that relations (\ref{eq: free_part_limit}) are the
first steps of recurrence relation
\begin{equation} \label{eq: frequency_Gaussian}
\Delta_{\nu+1}(k) = \frac{\nu+1}{\nu}\Delta_{\nu}(k),
\end{equation}
which obeys the Gaussian relaxation for the density
autocorrelation function
\begin{equation}
\phi_{coh}(k,t) = \mathrm{e}^{-(v_Tkt)^2/2}.
\end{equation}
Then, for the dynamic structure factor in the frequency domain one
has the single Gaussian-like function located at the zeroth
frequency. This result is completely reasonable, since it means
that the  short time dynamics in the range of high values of $k$
is defined by free particle movements, and the relaxation occurs
over a single time scale $\tau \sim \sqrt{1/\Delta_1(k)} \sim
(v_Tk)^{-1}$.

\subsection{Microscopic dynamics at wave numbers $k\leq k_m$}

\noindent At the condition for wave numbers $k \leq k_m$, the
corresponding spatial ranges of a system can be filled by few
particles only. Within such the ranges, the description of the
proper dynamics is relevant if it is performed in terms of two-,
three- and four-particle distribution functions, which are
contained in the first four frequency parameters, $\Delta_{\nu}$,
$\nu=1,\ 2,\ 3,\ 4$.

Moreover, the treatment of experimental $I(k,\omega)$-data of
inelastic X-ray scattering \cite{Mokshin_JPCM_2003} as well as the
numerical molecular dynamics simulations results for liquid alkali
metals near melting
\cite{Mokshin_JPCM_2006,Mokshin_JETP_2009,Mokshin_JETP_2006}
indicate that there is the correspondence for this range of wave
number:
\begin{equation} \label{eq: equality}
\Delta_4(k) \simeq \Delta_5(k).
\end{equation}
The extension of this equality to the frequency parameters of
higher order
\begin{equation} \label{eq: equality_extension}
\Delta_{\nu}=\Delta_{4}, \ \ \ \nu > 4,
\end{equation}
allows one to define the term $\tilde{\phi}_3(k,s)$ of the
chain~(\ref{eq: freq_chain}) by analogy with the model
case~(\ref{eq: condition_for_Bessel})
\begin{equation} \label{eq: closure}
\tilde{\phi}_3(k,s)=\frac{-s+\sqrt{s^2+4\Delta_4(k)}}{2\Delta_4(k)}.
\end{equation}
It is necessary to note that extension~(\ref{eq:
equality_extension}) has a clear sense at consideration of the
transition into the regime of high values of $k$.
Equation~(\ref{eq: frequency_Gaussian}) indicates on the equality
$\Delta_{\nu}(k)=\Delta_{\nu+1}(k)$ at limit of high $\nu$-values,
i.e. $\lim_{\nu \rightarrow \infty}
\Delta_{\nu}(k)/\Delta_{\nu+1}(k)=1$. Thus, for the regime of $k
\leq k_m$ the equality of frequency parameters arises at lower
level of chain~(\ref{eq: chain}) and smaller values of $\nu$.

Then, going down over fraction~(\ref{eq: continued_fraction}) to
the term $\tilde{\phi}_2(k,s)$ one obtains
\begin{eqnarray}
\tilde{\phi}_2(k,s)=\frac{2\Delta_4(k)}{s[2\Delta_4(k)-
\Delta_3(k)] +\Delta_3(k)\sqrt{s^2+ 4\Delta_4(k)}}. \nonumber \\
\label{eq: phi2}
\end{eqnarray}
Then, the dynamic structure factor will be as the next
\begin{eqnarray}
\label{eq: dsf_M3M4} S(k, \omega)&=& \frac{S(k)}{2 \pi}
\Delta_{1}(k) \Delta_{2}(k) \Delta_{3}(k) \sqrt{4 \Delta_{4}(k)-
\omega^{2}} \left \lbrace \Delta_{1}^{2}(k) \Delta_{3}^{2}(k)
\right.
                                                       \nonumber \\
&+&\omega^{2} \left [\Delta_{1}^{2}(k) \Delta_{4}(k)-2
\Delta_{1}(k) \Delta_{3}^{2}(k) - \Delta_{1}^{2}(k) \Delta_{3}(k)
+2 \Delta_{1}(k) \Delta_{2}(k) \Delta_{4}(k) - \Delta_{1}(k)
\Delta_{2}(k) \Delta_{3}(k) \right. \nonumber\\
& & \ \ \ \ \ \ \ \ \ \ \ \ \  + \left .\Delta_{2}^{2}(k)
\Delta_{4}(k) \right ]
                                                     \nonumber \\
&+&\omega^{4}\left [ \Delta_{3}^{2}(k)-2 \Delta_{1}(k)
\Delta_{4}(k) + 2 \Delta_{1}(k) \Delta_{3}(k) - 2 \Delta_{2}(k)
\Delta_{4}(k) + \Delta_{2}(k) \Delta_{3}(k)\right]
                                                       \nonumber\\
&+& \left . \omega^{6}\left [ \Delta_{4}(k) - \Delta_{3}(k) \right
] \right \rbrace^{-1}.
\end{eqnarray}
It is necessary to note that the dynamic structure factor
$S(k,\omega)$ of the form (\ref{eq: dsf_M3M4}) reproduces the
three-peak structure in the frequency domain (at the fixed $k$),
which is observable within the experiments of inelastic neutron
scattering and inelastic x-ray scattering. This spectral form is
very similar to the known Rayleigh -Mandelshtam-Brillouin triplet
at light scattering corresponded to the hydrodynamic limit ($t
\rightarrow \infty$ and $k \rightarrow 0$). Here, one peak of
$S(k,\omega)$ is located at the zeroth frequency ($\omega=0$),
while two other peaks -- the so-called doublet -- are located at
the finite frequencies ($\pm \omega_{L}$).

\subsection{Relation with hydrodynamics}

\noindent The features of the high-frequency (inelastic) peaks of
the dynamic structure factor $S(k,\omega)$ will be defined by the
solution of the next equation
\begin{equation} \label{eq: disp_eq_first}
s + \frac{\Delta_1(k)}{s} + \Delta_2(k) \tilde{\phi}_2(k,s) = 0,
\end{equation}
which is general within the recurrence relation
approach~\cite{Mokshin_JPCM_2006}. In the considered case, the
function $\tilde{\phi}_2(k,s)$ for the presented scheme has the
form~(\ref{eq: phi2}), while equation~(\ref{eq: disp_eq_first})
will have complex solutions $s= \Re[s(k)]+\mathrm{i}\Im[s(k)]$.
Here, the imaginary part $\Im[s(k)]$ defines the high-frequency
peak positions, while the real part $\Re[s(k)]$ is associated with
the widths of the peaks.

To analyze equation~(\ref{eq: disp_eq_first}) we introduce two
dimensionless quantities (at the fixed $k$):
\begin{subequations}
\begin{equation} \label{eq: xi}
\xi(k)=\frac{s^2}{\Delta_4(k)},
\end{equation}
\begin{equation} \label{eq: varrho}
\varsigma(k) = \frac{2\Delta_4(k)}{\Delta_3(k)}-1.
\end{equation}
\end{subequations}
At the transition to the hydrodynamical limit the next condition
should be satisfied:
\[
|\xi(k)|\ll 1,
\]
that means the consideration of the large time scales (small
frequencies). Then, equation~(\ref{eq: disp_eq_first}) takes the
following form:
\begin{equation} \label{eq: disp_eq_hydro}
s^3 + \frac{2\sqrt{\Delta_4(k)}}{\varsigma(k)}s^2 + \left [
\Delta_1(k) + \Delta_2(k) + \frac{\Delta_2(k)}{\varsigma(k)}
\right ]s + \frac{2 \Delta_1(k)\sqrt{\Delta_4(k)}}{\varsigma(k)} =
0.
\end{equation}
The solution of this cubic equation can be found within the
convergent scheme for approximating solutions applied by Mountain
(see reference~\cite{Mountain_RMP_1966})
\begin{subequations}
\begin{equation}
s_{1,2}(k) = \pm \sqrt{\Delta_1(k) \left \{ 1+
\frac{\Delta_2(k)[1+\varsigma(k)]}{\Delta_1(k)\varsigma(k)} \right
\}} - \frac{\sqrt{\Delta_4(k)}}{\varsigma(k)}\left ( 1 -
\frac{\Delta_1(k)\varsigma(k)}{\Delta_1(k)\varsigma(k)+\Delta_2(k)\varsigma(k)
+\Delta_2(k)} \right ),
\end{equation}
\begin{equation}
s_3(k) =  - \frac{2 \Delta_4(k)}{\varsigma(k)}
\frac{\Delta_1(k)\varsigma(k)}{\Delta_1(k)\varsigma(k)+\Delta_2(k)\varsigma(k)
+\Delta_2(k)}.
\end{equation}
\end{subequations}
These approximated solutions corresponds to the results of the
hydrodynamical Landau-Placzek theory with the following
parameters: the adiabatic sound velocity $c_s$ is
\[
(c_s k)^2 = \lim_{k \rightarrow 0}(v_T k)^2 = \lim_{k \rightarrow
0} \Delta_1(k),
\]
the sound damping parameter is
\[
\Gamma = \frac{\gamma - 1}{\gamma}
\frac{\sqrt{\Delta_4(k)}}{\varsigma(k)},
\]
where the ratio of the specific heat at constant pressure to the
specific heat at constant volume $\gamma = c_p/c_v$ is
\[
\gamma = 1 + \frac{\Delta_2(k) +
\Delta_2(k)\varsigma(k)}{\Delta_1(k)\varsigma(k)}.
\]
Thus, the expression for the dynamic structure factor
$S(k,\omega)$ of the form~(\ref{eq: dsf_M3M4}) satisfies
completely the transition into hydrodynamic regime; and the
parameters of hydrodynamic theory are expressed through the
frequency parameters $\Delta_1(k)$, $\Delta_2(k)$, $\Delta_3(k)$
and $\Delta_4(k)$.

\section{Conclusions}

\noindent In this paper, we have presented the approach, by means
of which the relaxation processes in many particle systems can be
studied. This approach reformulates the equation of motion for the
considered dynamical variable in terms of the recurrence
relations, where the set of frequency parameters appear.  In fact,
the recurrence relations approach states that if the frequency
parameters or the correspondence between of them are known, then
the solution for the TACF can be found. Moreover, it imposes the
general routine to define these frequency parameters; however, the
specific analytical expressions depend on a concrete problem
studied~\cite{Wierling_PRE_2010,Wierling_EPJB_2012,Wierling_CCP_2012,Lee_Physica_Scr_1987}.

Here, we have shown some simple model cases with the exact
solutions, which are realized, nevertheless, in the physical
problems. As the other nontrivial case, we have considered the
problem of the description of local density fluctuations in simple
liquids. As it was shown, the solution for the spectral image of
the local density fluctuation TACF can be obtained within the
recurrence relations, and this solution is correctly consistent
with the known asymptotic scenarios.

The important advantage of the approach is the absence of the
abstract parameters at the construction of theoretical
description. All the quantities are expressed only through the
frequency parameters, which contain the details of interparticle
interactions and static structure correlations. Therefore, such
the approach,  in our opinion, is the perfect one within of which
it is possible to construct the explanation how the many particle
system structure defines the system dynamics.

\section{Acknowledgements}
\noindent  We would like to thank M. Howard Lee (University of
Georgia, USA) for very useful discussions.

\end{document}